\documentclass[10pt, conference]{IEEEtran}
\IEEEoverridecommandlockouts % enables \thanks (funding note) in conference mode
\usepackage{amsmath}
\usepackage{amssymb}
\usepackage{amsthm}
\usepackage{algorithm}
\usepackage{algorithmic}
\usepackage{bm}
\usepackage{graphicx}
\usepackage[caption=false,font=footnotesize]{subfig}
\usepackage{cite}
\usepackage{xcolor}
\usepackage{booktabs}
\newtheorem{proposition}{Proposition}
\newtheorem{remark}{Remark}

\begin{document}

\title{
A Faster and Efficient Indoor User Localization using Bistatic MIMO Radar
\thanks{The work of Neel Kanth Kundu was supported in part by the DST INSPIRE Faculty Fellowship (Reg. No.: IFA22-ENG 344), ANRF Prime Minister Early Career Research Grant (ANRF/ECRG/2024/000324/ENS), and IIT Delhi's New Faculty Seed Grant (MI02855G).}
}
\author{
    \IEEEauthorblockN{
        Shadab Anjum\IEEEauthorrefmark{1},
        Sourav Banerjee\IEEEauthorrefmark{1}, and
        Neel Kanth Kundu\IEEEauthorrefmark{1}\IEEEauthorrefmark{2}
    }\\
    \IEEEauthorblockA{\small
        \IEEEauthorrefmark{1}Centre for Applied Research in Electronics (CARE),
        Indian Institute of Technology Delhi, New Delhi, India
    }
    \IEEEauthorblockA{\small
        \IEEEauthorrefmark{2}Bharti School of Telecommunication Technology and Management,
        Indian Institute of Technology Delhi, New Delhi, India\\
        \small Email: shadabanjum022@gmail.com, souravbanerjer9@gmail.com, and neelkanth@iitd.ac.in
    }
}
\maketitle

%% ====================================================b
\begin{abstract}
Indoor user localization is essential for applications such as asset tracking, robotic navigation, and smart-building automation where satellite-based positioning is unavailable. Bistatic MIMO radar exploits bistatic range measurements and geometric diversity from distributed antennas, but low-SNR conditions make the resulting estimation ill-conditioned. We formulate indoor localization as a Tikhonov-regularized nonlinear least-squares problem and solve it with a damped Gauss--Newton (DGN) algorithm, in which each iteration reduces to a closed-form $3\times3$ linear solve followed by an Armijo backtracking line search. Simulations show accuracy on par with, and runtime far faster than, an SDP-based benchmark.
\end{abstract}


\section{Introduction}

Indoor target localization i.e. estimating the position of a person, robot, or asset within an enclosed environment is important for navigation, surveillance, and automation where satellite-based positioning is unavailable \cite{richards2005fundamentals}. Radar-based localization addresses this by processing reflected signals without requiring cooperative devices on the target~\cite{Wang2019}.

In recent years, multiple-input multiple-output (MIMO) radar has emerged as a powerful paradigm that significantly enhances localization performance compared to conventional phased-array systems~\cite{Haimovich_2008}. By employing multiple transmit and receive antennas, MIMO radar systems exploit waveform diversity and spatial diversity, leading to improved parameter identifiability and robustness. Depending on the antenna configuration, MIMO radar systems can be broadly categorized into colocated and widely separated antenna systems. The latter, often referred to as distributed or bistatic/multistatic MIMO radar, provides substantial gains in localization accuracy due to geometric diversity and independent observation paths \cite{Guo2019}. Such distributed configurations are especially well suited to indoor settings, where antennas can be mounted around the perimeter of a room so that the target lies within their convex hull -- a geometric property that proves useful for both algorithm initialization and feasible-set design.

Target localization in distributed MIMO radar is commonly formulated using bistatic range measurements, which represent the sum of distances from transmitter-to-target and target-to-receiver ~\cite{beck_2008}. This indirect localization approach decouples signal processing from position estimation and has been widely studied due to its practical relevance. However, the resulting estimation problem is inherently nonlinear and nonconvex, and its conditioning worsens under realistic, low signal-to-noise ratio (SNR) indoor noise conditions. Maximum likelihood estimation (MLE) formulations are statistically optimal but computationally expensive and sensitive to initialization, while least-squares (LS) based methods offer reduced complexity at the expense of degraded performance under high noise \cite{Amiri2017}; Tikhonov regularization can mitigate this ill-conditioning, at the cost of introducing an additional design parameter~\cite{Golub1999TikhonovRA}.

To address these challenges, semidefinite programming (SDP) relaxation techniques have been proposed to convexify the localization problem and obtain tractable solutions. In particular, the work in \cite{Wang2019} formulates the bistatic MIMO radar localization problem as a nonconvex optimization and applies SDP relaxation to obtain a convex approximation. Furthermore, a penalty term is introduced to improve solution quality and mitigate ambiguity in the lifted variables. While SDP-based methods provide good accuracy and robustness, they typically incur high computational complexity due to interior-point solvers, making them less suitable for large-scale, low-latency indoor systems or real-time implementations. Additionally, the relaxation may not always be tight, especially under adverse noise conditions.

Motivated by these limitations, this paper develops a \emph{damped Gauss--Newton} (DGN) algorithm for indoor bistatic MIMO radar target localization.
The Gauss--Newton method solves a nonlinear least-squares problem by repeatedly linearizing the residual about the current iterate and minimizing the resulting convex quadratic model; its damped variant scales each Gauss--Newton step by a stepsize chosen through a line search, which guarantees monotonic decrease of the objective even when the linearization is inaccurate far from the solution~\cite{nocedal2006numerical}. Our approach retains the problem structure in its original nonlinear least-squares form and avoids lifting to higher-dimensional semidefinite variables, thereby significantly reducing computational burden: for the bistatic range model, each Gauss--Newton step reduces to a closed-form $3\times 3$ linear solve. The resulting DGN algorithm guarantees monotonic decrease of the objective and, as long as its iterates stay away from the antenna positions, every limit point of the iterates is a stationary point (Proposition~\ref{prop:convergence}).

The key contributions of this work are summarized as follows:
\begin{enumerate}
    \item A rigorous nonlinear least-squares formulation of the MIMO
          radar localization problem with Tikhonov regularization.
    \item A Gauss--Newton update derived from first-order Taylor
          linearization of the bistatic range vector and its Jacobian,
          which reduces each iteration to a closed-form $3\times 3$
          linear solve.
    \item Integration of an Armijo backtracking line search on the
          Gauss--Newton direction to ensure descent when the
          linearization radius is large.
\end{enumerate}
The remainder of the paper is organized as follows. Section~\ref{sec:model}
presents the system model and problem formulation. Section~\ref{sec:sca}
derives the damped Gauss--Newton algorithm in detail. Section~\ref{sec:sim} presents
numerical results, and Section~\ref{sec:conclusion} concludes the paper.
%% =====================================================================
%% =====================================================================

\begin{figure}[t]
    \centering
    \includegraphics[width=1.0\linewidth]{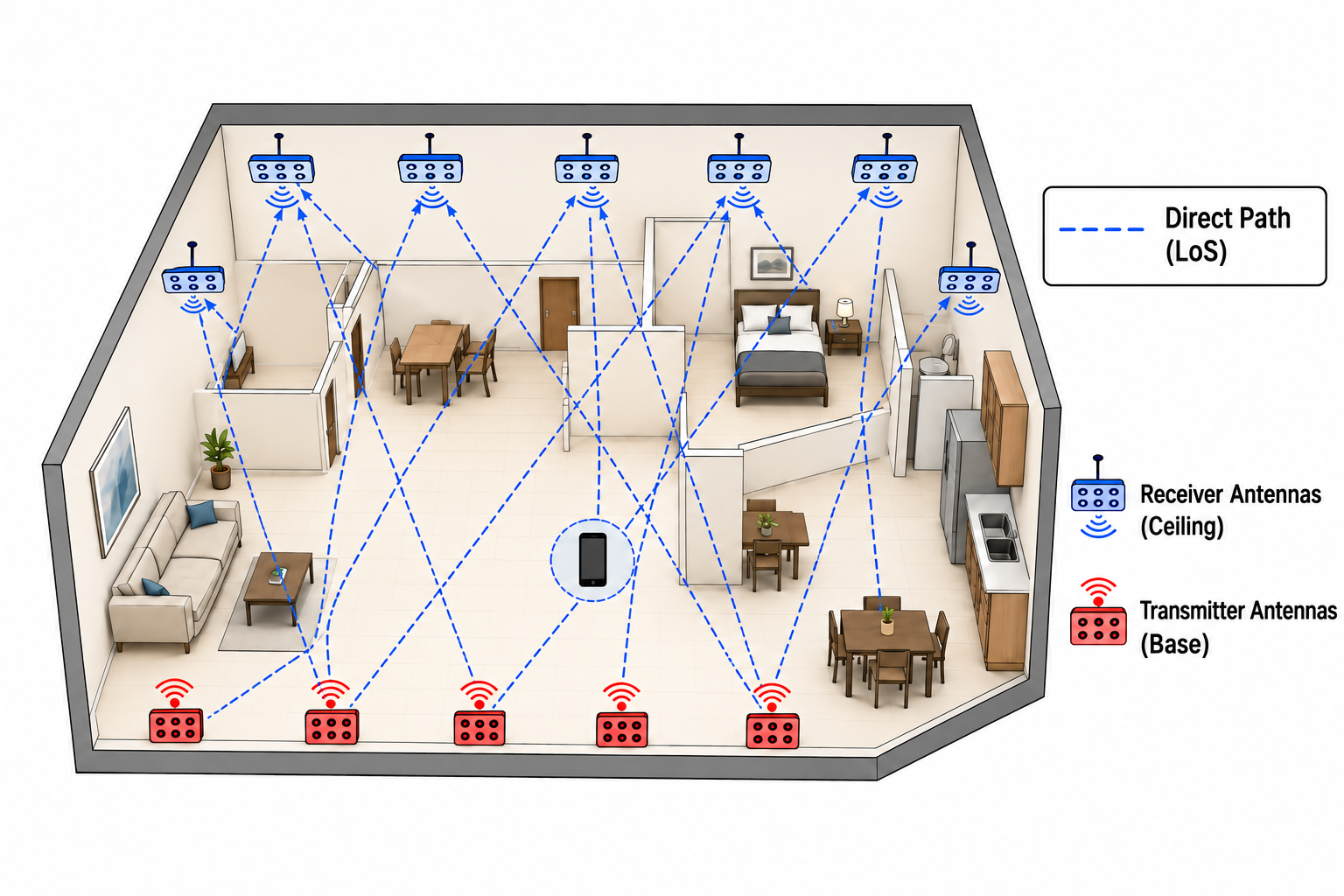}
    \caption{Bistatic MIMO radar-based indoor user localization.}
    \label{fig:schematic_diagram}
\end{figure}

\section{System Model and Problem Formulation}
\label{sec:model}

\subsection{MIMO Radar Geometry}

Consider a MIMO radar system~\cite{Bekkerman2006TargetDA} comprising $M$ transmitters located at
positions $\{\mathbf{x}_m^t\}_{m=1}^{M} \subset \mathbb{R}^3$ and
$N$ receivers located at positions $\{\mathbf{x}_n^r\}_{n=1}^{N}
\subset \mathbb{R}^3$. A point target is located at an unknown
position $\mathbf{u} \in \mathbb{R}^3$. The bistatic range associated with the $m$-th transmitter and $n$-th receiver pair is defined as
the sum of the transmit range and receive range:
\begin{equation}
    \rho_{mn}(\mathbf{u}) \;=\;
    \underbrace{\|\mathbf{u} - \mathbf{x}_m^t\|}_{R_m^t(\mathbf{u})}
    \;+\;
    \underbrace{\|\mathbf{u} - \mathbf{x}_n^r\|}_{R_n^r(\mathbf{u})}
    \label{eq:bistatic_range}
\end{equation}
where $\|\cdot\|$ denotes the Euclidean norm. A schematic diagram of bistatic MIMO radar based indoor user localization is shown in Fig.~\ref{fig:schematic_diagram}.

\subsection{Measurement Model}

The measured bistatic range for the $(m,n)$ pair is modeled as:
\begin{equation}
    r_{mn} \;=\; \rho_{mn}(\mathbf{u}) + \varepsilon_{mn}
    \;=\; R_m^t(\mathbf{u}) + R_n^r(\mathbf{u}) + \varepsilon_{mn}
    \label{eq:meas_model_scalar}
\end{equation}
where $\varepsilon_{mn} \sim \mathcal{N}(0,\,\sigma^2)$ is additive
white Gaussian noise (AWGN), independent across all $(m,n)$ pairs.

\subsection{Vectorized Measurement Model}

Define the range vector $\mathbf{g}(\mathbf{u}) \in \mathbb{R}^{M+N}$
as the concatenation of all transmit and receive ranges:
\begin{equation}
    \mathbf{g}(\mathbf{u}) \;=\;
    \begin{bmatrix}
        R_1^t(\mathbf{u}) \;\cdots\; R_M^t(\mathbf{u}) \;
        R_1^r(\mathbf{u}) \;\cdots\; R_N^r(\mathbf{u})
    \end{bmatrix}^\top
    \label{eq:range_vector}
\end{equation}

Stack all $MN$ measurements in lexicographic order ($n$ varies
fastest) to form $\mathbf{r} \in \mathbb{R}^{MN}$. The full
measurement model then takes the compact linear-in-$\mathbf{g}$ form:
\begin{equation}
    \mathbf{r} \;=\; \mathbf{D}\,\mathbf{g}(\mathbf{u}) + \boldsymbol{\varepsilon}
    \label{eq:meas_model_vector}
\end{equation}
where the mixing matrix $\mathbf{D} \in \mathbb{R}^{MN \times (M+N)}$
is defined as:
\begin{equation}
    \mathbf{D} \;=\;
    \left[\,\mathbf{I}_M \otimes \mathbf{1}_N,\quad
           \mathbf{1}_M \otimes \mathbf{I}_N\,\right]
    \label{eq:D_matrix}
\end{equation}
with $\otimes$ denoting the Kronecker product, $\mathbf{I}_k$ the
$k\times k$ identity matrix, and $\mathbf{1}_k$ the all-ones column
vector of length $k$. The $(m-1)N+n$-th row of $\mathbf{D}$ contains
a $1$ in column $m$ and a $1$ in column $M+n$, encoding the
bistatic additive structure of \eqref{eq:meas_model_scalar}.

\subsection{Optimization Problem}

The maximum-likelihood (ML) estimator under AWGN is
obtained by minimizing the sum of squared residuals. To improve numerical
stability and mitigate ill-conditioning caused by near-degenerate sensor
geometries, a Tikhonov regularization term is incorporated,
leading to the following optimization problem:
\begin{equation}
    \min_{\mathbf{u} \in \mathbb{R}^3}
    F(\mathbf{u})
    =
    \underbrace{\left\|\mathbf{r}
    - \mathbf{D}\mathbf{g}(\mathbf{u})\right\|^2}_{\text{data fidelity}}
    +
    \underbrace{\eta\left\|\mathbf{g}(\mathbf{u})\right\|^2}_{\text{Tikhonov regularization}},
    \label{eq:objective}
\end{equation}
where $\eta>0$ is the regularization parameter controlling the trade-off
between data fidelity and regularization~\cite{5714759}. Although each component of
$\mathbf{g}(\mathbf{u})$ is a convex Euclidean distance function, the mapping
$\mathbf{g}(\mathbf{u})$ is nonlinear. Consequently, the data-fidelity term
$\|\mathbf{r}-\mathbf{D}\mathbf{g}(\mathbf{u})\|^2$ is generally non-convex,
making the overall objective function $F(\mathbf{u})$ non-convex. Therefore,
obtaining the global optimum is challenging. However, since $F(\mathbf{u})$
is a sum of squared residuals, problem \eqref{eq:objective} is a nonlinear
least-squares problem, which motivates the Gauss--Newton approach developed
in the following section.

%% =====================================================================
\section{Damped Gauss--Newton Algorithm}
\label{sec:sca}

Stacking the two terms of \eqref{eq:objective}, the objective can be
written as a single sum of squares,
\begin{equation}
    F(\mathbf{u}) \;=\; \left\|\boldsymbol{\psi}(\mathbf{u})\right\|^2,
    \qquad
    \boldsymbol{\psi}(\mathbf{u}) \;=\;
    \begin{bmatrix}
        \mathbf{r} - \mathbf{D}\mathbf{g}(\mathbf{u})\\
        \sqrt{\eta}\,\mathbf{g}(\mathbf{u})
    \end{bmatrix},
    \label{eq:stacked_residual}
\end{equation}
so that \eqref{eq:objective} is a standard nonlinear least-squares
problem. The Gauss--Newton method solves such problems by linearizing
$\boldsymbol{\psi}(\mathbf{u})$ about the current iterate
$\mathbf{u}^{(k)}$ and minimizing the resulting convex quadratic model
$\tilde{F}(\mathbf{u}\mid\mathbf{u}^{(k)})$ to obtain a candidate
\begin{equation}
    \hat{\mathbf{u}}(\mathbf{u}^{(k)}) \;=\;
    \arg\min_{\mathbf{u}\in\mathbb{R}^3} \tilde{F}(\mathbf{u}\mid\mathbf{u}^{(k)}).
    \label{eq:sca_subproblem}
\end{equation}
In the damped Gauss--Newton method, the next iterate moves only a
fraction of the way toward this candidate,
\begin{equation}
    \mathbf{u}^{(k+1)} \;=\; \mathbf{u}^{(k)}
    + \gamma^{(k)}\left(\hat{\mathbf{u}}(\mathbf{u}^{(k)}) - \mathbf{u}^{(k)}\right),
    \label{eq:sca_smoothing}
\end{equation}
where the stepsize $\gamma^{(k)}\in(0,1]$ is chosen by a line
search~\cite{nocedal2006numerical}. The quadratic model need not
upper-bound $F$; monotonic descent is instead guaranteed by the line
search, detailed in Section~\ref{sec:armijo}.

\subsection{Jacobian of the Range Vector}

Constructing the Gauss--Newton model requires the Jacobian
$\mathbf{J}(\mathbf{u}) \in \mathbb{R}^{(M+N)\times 3}$ of
$\mathbf{g}(\mathbf{u})$, whose $m$-th and $(M+n)$-th rows are the
unit direction vectors from the $m$-th transmitter and $n$-th
receiver to the target, respectively:
\begin{align}
    \left[\mathbf{J}(\mathbf{u})\right]_m
    &\;=\; \frac{\left(\mathbf{u} - \mathbf{x}_m^t\right)^\top}
                {\left\|\mathbf{u} - \mathbf{x}_m^t\right\|}
    \label{eq:jac_t}\\
    \left[\mathbf{J}(\mathbf{u})\right]_{M+n}
    &\;=\; \frac{\left(\mathbf{u} - \mathbf{x}_n^r\right)^\top}
                {\left\|\mathbf{u} - \mathbf{x}_n^r\right\|}
    \label{eq:jac_r}
\end{align}
which are well defined whenever $\mathbf{u} \neq \mathbf{x}_m^t$ and
$\mathbf{u} \neq \mathbf{x}_n^r$ for all $m,n$.
\begin{algorithm}[t]
\caption{Damped Gauss--Newton for MIMO radar target localization}
\label{alg:SCA-BCD}
\begin{algorithmic}[1]
\REQUIRE Measurements $\mathbf{r}$, mixing matrix $\mathbf{D}$,
         sensor positions $\{\mathbf{x}_m^t\}$, $\{\mathbf{x}_n^r\}$,
         regularization weight $\eta > 0$, tolerance $\tau > 0$,
         backtracking stepsize parameters $\beta = 0.5$, $c = 10^{-4}$,
         max iterations $K_{\max}$

\ENSURE Estimated target position $\hat{\mathbf{u}}$

\STATE Initialize $\mathbf{u}^{(0)}$ via \eqref{eq:init}

\FOR{$k = 0, 1, 2, \ldots, K_{\max}-1$}

    %\STATE \textit{--- SCA surrogate minimization ---}
    \STATE Compute $\mathbf{g}^{(k)} = \mathbf{g}(\mathbf{u}^{(k)})$ and
           $\mathbf{J}^{(k)} = \mathbf{J}(\mathbf{u}^{(k)})$
           via \eqref{eq:range_vector}, \eqref{eq:jac_t}, \eqref{eq:jac_r}
    \STATE Form $\mathbf{A}$, $\mathbf{b}$, $\mathbf{A}_e$, $\mathbf{b}_e$
           via \eqref{eq:Ab}--\eqref{eq:Aebe} 
           % (strongly convex surrogate
           % $\tilde{F}(\cdot\mid\mathbf{u}^{(k)})$)
    \STATE Compute $\mathbf{H} = \mathbf{A}^\top\mathbf{A}
           + \eta\,\mathbf{A}_e^\top\mathbf{A}_e$ and
           $\mathbf{q} = \mathbf{A}^\top\mathbf{b}
           - \eta\,\mathbf{A}_e^\top\mathbf{b}_e$
           via \eqref{eq:Hq}
    \STATE Solve $\hat{\mathbf{u}}(\mathbf{u}^{(k)}) = \mathbf{H}^{-1}\mathbf{q}$
    \STATE Set Gauss--Newton direction
           $\mathbf{d}^{(k)} = \hat{\mathbf{u}}(\mathbf{u}^{(k)}) - \mathbf{u}^{(k)}$

    %\STATE \textit{--- Backtracking stepsize (smoothing) ---}
    \STATE Compute residual
           $\boldsymbol{\delta}^{(k)} = \mathbf{r} - \mathbf{D}\mathbf{g}^{(k)}$
           and exact gradient
           $\nabla F(\mathbf{u}^{(k)}) = -2[\mathbf{J}^{(k)}]^\top
           \mathbf{D}^\top\boldsymbol{\delta}^{(k)}
           + 2\eta[\mathbf{J}^{(k)}]^\top\mathbf{g}^{(k)}$
           via \eqref{eq:exact_grad_1}
    \STATE Set $s \leftarrow 0$
    \WHILE{$F\!\left(\mathbf{u}^{(k)} + \beta^s \mathbf{d}^{(k)}\right)
            > F\!\left(\mathbf{u}^{(k)}\right)
            + c\,\beta^s\,[\nabla F(\mathbf{u}^{(k)})]^\top
            \mathbf{d}^{(k)}$}
        \STATE $s \leftarrow s + 1$
    \ENDWHILE
    \STATE Set $\gamma^{(k)} = \beta^s$
    \STATE $\mathbf{u}^{(k+1/2)} \leftarrow
           \mathbf{u}^{(k)} + \gamma^{(k)}\,\mathbf{d}^{(k)}$
           \hfill 
           %\COMMENT{SCA smoothing step \eqref{eq:sca_smoothing}}

    %\STATE \textit{--- Intra-iteration BCD refinement ---}
    \STATE Compute $\mathbf{g}^{(k+1/2)} =
           \mathbf{g}(\mathbf{u}^{(k+1/2)})$ and
           $\mathbf{J}^{(k+1/2)} = \mathbf{J}(\mathbf{u}^{(k+1/2)})$
    \STATE Form re-linearized model matrices
           $\mathbf{A}'$, $\mathbf{b}'$, $\mathbf{A}_e'$, $\mathbf{b}_e'$
           via \eqref{eq:Ab}--\eqref{eq:Aebe} at $\mathbf{u}^{(k+1/2)}$;
           solve $\tilde{\mathbf{u}} =
           (\mathbf{A}'^\top\mathbf{A}' +
           \eta\,\mathbf{A}_e'^\top\mathbf{A}_e')^{-1}
           (\mathbf{A}'^\top\mathbf{b}' -
           \eta\,\mathbf{A}_e'^\top\mathbf{b}_e')$
    \STATE \textbf{if} $F(\tilde{\mathbf{u}}) < F(\mathbf{u}^{(k+1/2)})$
           \textbf{then}
           $\mathbf{u}^{(k+1)} \leftarrow \tilde{\mathbf{u}}$
           \textbf{else}
           $\mathbf{u}^{(k+1)} \leftarrow \mathbf{u}^{(k+1/2)}$

    \IF{$\dfrac{\|\mathbf{u}^{(k+1)} - \mathbf{u}^{(k)}\|}
              {\max(\|\mathbf{u}^{(k)}\|,\,\epsilon_0)} < \tau$}
        \STATE \textbf{break}
    \ENDIF

\ENDFOR

\RETURN $\hat{\mathbf{u}} = \mathbf{u}^{(k+1)}$
\end{algorithmic}
\end{algorithm}
\subsection{Gauss--Newton Linearization}

Linearizing $\mathbf{g}(\mathbf{u})$ about $\mathbf{u}^{(k)}$,
$\mathbf{g}(\mathbf{u}) \approx \mathbf{g}^{(k)} +
\mathbf{J}^{(k)}(\mathbf{u}-\mathbf{u}^{(k)})$ with
$\mathbf{g}^{(k)}=\mathbf{g}(\mathbf{u}^{(k)})$,
$\mathbf{J}^{(k)}=\mathbf{J}(\mathbf{u}^{(k)})$, and substituting into
\eqref{eq:objective} yields the Gauss--Newton quadratic model
\begin{equation}
    \tilde{F}\!\left(\mathbf{u} \mid \mathbf{u}^{(k)}\right) \;=\;
    \left\|\mathbf{b} - \mathbf{A}\,\mathbf{u}\right\|^2
    \;+\; \eta\,\left\|\mathbf{b}_e + \mathbf{A}_e\,\mathbf{u}\right\|^2,
    \label{eq:surrogate}
\end{equation}
\begin{align}
    \mathbf{A}   &\;=\; \mathbf{D}\mathbf{J}^{(k)},
    \qquad
    \mathbf{b}   \;=\; \mathbf{r} - \mathbf{D}\mathbf{g}^{(k)}
                  + \mathbf{A}\,\mathbf{u}^{(k)}
    \label{eq:Ab}\\
    \mathbf{A}_e &\;=\; \mathbf{J}^{(k)},
    \qquad
    \mathbf{b}_e \;=\; \mathbf{g}^{(k)} - \mathbf{J}^{(k)}\mathbf{u}^{(k)}.
    \label{eq:Aebe}
\end{align}

\subsection{Closed-Form Gauss--Newton Step}

Setting $\nabla_{\mathbf{u}}\tilde{F}=\mathbf{0}$ gives the
$3\times 3$ normal equation $\mathbf{H}\,\hat{\mathbf{u}}(\mathbf{u}^{(k)}) = \mathbf{q}$,
with
\begin{equation}
    \mathbf{H} \;=\; \mathbf{A}^\top\mathbf{A}
                \;+\; \eta\,\mathbf{A}_e^\top\mathbf{A}_e,
    \qquad
    \mathbf{q} \;=\; \mathbf{A}^\top\mathbf{b}
                \;-\; \eta\,\mathbf{A}_e^\top\mathbf{b}_e.
    \label{eq:Hq}
\end{equation}
Note that $\mathbf{H} = \left[\mathbf{J}^{(k)}\right]^\top
(\mathbf{D}^\top\mathbf{D} + \eta\,\mathbf{I})\,\mathbf{J}^{(k)}$.
Since $\mathbf{D}$ has the null vector
$[\mathbf{1}_M^\top,\,-\mathbf{1}_N^\top]^\top$, the term $\eta>0$
ensures $\mathbf{D}^\top\mathbf{D} + \eta\,\mathbf{I} \succ \mathbf{0}$;
hence $\mathbf{H}$ is symmetric positive definite whenever
$\mathbf{J}^{(k)}$ has full column rank, i.e., whenever the unit
direction vectors in \eqref{eq:jac_t}--\eqref{eq:jac_r} span
$\mathbb{R}^3$, which holds at every $\mathbf{u}$ that is not a sensor
position whenever the $M+N$ sensors are not coplanar (see the proof of
Proposition~\ref{prop:convergence}).
The minimizer of \eqref{eq:sca_subproblem} is then available in
closed form:
\begin{equation}
    \hat{\mathbf{u}}(\mathbf{u}^{(k)}) \;=\; \mathbf{H}^{-1}\mathbf{q}.
    \label{eq:mm_update}
\end{equation}

\begin{remark}[Gauss--Newton interpretation]
The Jacobian of $\boldsymbol{\psi}$ in \eqref{eq:stacked_residual} is
$\mathbf{J}_\psi = [-(\mathbf{D}\mathbf{J})^\top,\ \sqrt{\eta}\,\mathbf{J}^\top]^\top$.
Hence the model \eqref{eq:surrogate} equals
$\|\boldsymbol{\psi}(\mathbf{u}^{(k)}) +
\mathbf{J}_\psi^{(k)}(\mathbf{u}-\mathbf{u}^{(k)})\|^2$,
$\mathbf{H} = [\mathbf{J}_\psi^{(k)}]^\top\mathbf{J}_\psi^{(k)}$, and
\begin{equation*}
    \mathbf{d}^{(k)} = \hat{\mathbf{u}}(\mathbf{u}^{(k)}) - \mathbf{u}^{(k)}
    = -\big([\mathbf{J}_\psi^{(k)}]^\top\mathbf{J}_\psi^{(k)}\big)^{-1}
    [\mathbf{J}_\psi^{(k)}]^\top\boldsymbol{\psi}(\mathbf{u}^{(k)}),
\end{equation*}
which is exactly the Gauss--Newton direction for
\eqref{eq:stacked_residual}.
\end{remark}

\subsection{Stepsize Rule: Backtracking Line Search}
\label{sec:armijo}
Since the linearization is only locally accurate, the full step
$\gamma^{(k)}=1$ need not decrease $F$; we therefore adopt a
backtracking line search. With Gauss--Newton direction
$\mathbf{d}^{(k)} = \hat{\mathbf{u}}(\mathbf{u}^{(k)}) - \mathbf{u}^{(k)}$
and exact gradient
\begin{align}
    \nabla_{\mathbf{u}} F\!\left(\mathbf{u}^{(k)}\right)
    &\;=\;
    -2\left[\mathbf{J}^{(k)}\right]^\top\!\mathbf{D}^\top
    \boldsymbol{\delta}^{(k)}
    \;+\; 2\eta\left[\mathbf{J}^{(k)}\right]^\top\mathbf{g}^{(k)}, \label{eq:exact_grad_1}\\
    \boldsymbol{\delta}^{(k)} &= \mathbf{r} - \mathbf{D}\mathbf{g}^{(k)}, \label{eq:exact_grad_2}
\end{align}
the stepsize is set to $\gamma^{(k)} = \beta^{s^*}$ where $\beta=0.5$ and
$s^*$ is the smallest non-negative integer satisfying the Armijo
condition~\cite{armijo1966minimization}
\begin{equation}
    F\!\left(\mathbf{u}^{(k)} + \beta^s\,\mathbf{d}^{(k)}\right)
    \;\leq\;
    F\!\left(\mathbf{u}^{(k)}\right)
    + c\,\beta^s\,\left[\nabla_{\mathbf{u}} F\!\left(\mathbf{u}^{(k)}\right)
    \right]^\top\!\mathbf{d}^{(k)},
    \label{eq:armijo}
\end{equation}
with $c = 10^{-4}$, giving
$\mathbf{u}^{(k+1/2)} = \mathbf{u}^{(k)} + \gamma^{(k)}\,\mathbf{d}^{(k)}$.
From \eqref{eq:Hq}, \eqref{eq:mm_update} and \eqref{eq:exact_grad_1},
$\mathbf{H}\,\mathbf{d}^{(k)} = -\tfrac{1}{2}\nabla_{\mathbf{u}}F(\mathbf{u}^{(k)})$,
so that $[\nabla_{\mathbf{u}}F(\mathbf{u}^{(k)})]^\top\mathbf{d}^{(k)}
= -2\,[\mathbf{d}^{(k)}]^\top\mathbf{H}\,\mathbf{d}^{(k)} < 0$ whenever
$\mathbf{u}^{(k)}$ is not stationary. Hence $\mathbf{d}^{(k)}$ is a
strict descent direction of $F$, \eqref{eq:armijo} is always
satisfiable in finitely many halvings, and $\gamma^{(k)}>0$ is
guaranteed to exist. Scaling the Gauss--Newton step by $\gamma^{(k)}$
in this way is what makes the method a \emph{damped} Gauss--Newton
method~\cite{nocedal2006numerical}.
\subsection{Safeguarded Full Gauss--Newton Step}

After the damped step \eqref{eq:sca_smoothing}\footnote{With $\mathbf{u}^{(k+1/2)} = \mathbf{u}^{(k)} + \gamma^{(k)}\,\mathbf{d}^{(k)}$, see \eqref{eq:armijo}.}, a second Gauss--Newton step is taken within
the same iteration, using a model freshly re-linearized at
$\mathbf{u}^{(k+1/2)}$:
\begin{enumerate}
    \item Compute $\mathbf{g}^{(k+1/2)} = \mathbf{g}(\mathbf{u}^{(k+1/2)})$
          and $\mathbf{J}^{(k+1/2)} = \mathbf{J}(\mathbf{u}^{(k+1/2)})$.
    \item Form updated model matrices analogous to
          \eqref{eq:Ab}--\eqref{eq:Aebe} using $\mathbf{u}^{(k+1/2)}$
          as the linearization point.
    \item Solve the resulting $3\times 3$ system to obtain
          the candidate $\tilde{\mathbf{u}}$.
    \item Accept this full step only if it yields strict descent:
\end{enumerate}
\begin{equation}
    \mathbf{u}^{(k+1)} \;=\;
    \begin{cases}
        \tilde{\mathbf{u}} &
        \text{if } F(\tilde{\mathbf{u}}) < F(\mathbf{u}^{(k+1/2)})\\
        \mathbf{u}^{(k+1/2)} & \text{otherwise}
    \end{cases}
    \label{eq:bcd_accept}
\end{equation}
Because this full step is accepted only when it decreases $F$, the
monotonic decrease guaranteed by the line search is preserved.

The algorithm is initialized at the sensor centroid,
\begin{equation}
    \mathbf{u}^{(0)} \;=\; \frac{1}{M+N}
    \left(\sum_{m=1}^{M}\mathbf{x}_m^t
    + \sum_{n=1}^{N}\mathbf{x}_n^r\right),
    \label{eq:init}
\end{equation}
which lies within the convex hull of the sensor array, ensuring that
the Jacobian denominators in \eqref{eq:jac_t}--\eqref{eq:jac_r} are
strictly positive at $\mathbf{u}^{(0)}$.

\subsection{Convergence Criterion}

The algorithm terminates when the relative displacement of the
position estimate falls below a threshold $\tau$:
\begin{equation}
    \frac{\left\|\mathbf{u}^{(k+1)} - \mathbf{u}^{(k)}\right\|}
         {\max\!\left(\left\|\mathbf{u}^{(k)}\right\|,\,\epsilon_0\right)}
    \;<\; \tau
    \label{eq:convergence}
\end{equation}
where $\tau > 0$ is the relative-displacement tolerance and
$\epsilon_0 > 0$ prevents division by zero when $\mathbf{u}^{(k)}$
is near the origin.

The full damped Gauss--Newton procedure is summarized in Algorithm~\ref{alg:SCA-BCD}.

\subsection{Convergence Analysis}

The objective $F$ is smooth everywhere except at the sensor positions
$\mathbf{x}_m^t$ and $\mathbf{x}_n^r$, where the Jacobian
\eqref{eq:jac_t}--\eqref{eq:jac_r} is undefined.

\begin{proposition}\label{prop:convergence}
Assume that the $M+N$ sensors are not coplanar and that the iterates of
Algorithm~\ref{alg:SCA-BCD} stay at a distance of at least $\delta>0$
from every sensor. Then $\{F(\mathbf{u}^{(k)})\}$ is nonincreasing and
converges, and every limit point of $\{\mathbf{u}^{(k)}\}$ is a
stationary point of $F$.
\end{proposition}

\begin{proof}
Away from the sensors, the unit vectors in
\eqref{eq:jac_t}--\eqref{eq:jac_r} span $\mathbb{R}^3$ (otherwise all
sensors would lie in one plane), so $\mathbf{J}$ has full column rank
and $\mathbf{H}\succ\mathbf{0}$. Hence $\mathbf{d}^{(k)}$ is a descent
direction (Section~\ref{sec:armijo}), and \eqref{eq:armijo} together
with \eqref{eq:bcd_accept} gives
$F(\mathbf{u}^{(k+1)})\le F(\mathbf{u}^{(k)})$; since $F\ge 0$,
$\{F(\mathbf{u}^{(k)})\}$ converges. Moreover,
$F(\mathbf{u})\to\infty$ as $\|\mathbf{u}\|\to\infty$, so all iterates
lie in the compact set $\mathcal{K}$ of points $\mathbf{u}$ with
$F(\mathbf{u})\le F(\mathbf{u}^{(0)})$ and distance at least $\delta$
from every sensor. On $\mathcal{K}$, $F$ is smooth, so $\nabla F$ is
Lipschitz continuous, and the eigenvalues of $\mathbf{H}$ lie in an
interval $[\lambda,\Lambda]$ with $\lambda>0$. Therefore
$\mathbf{d}^{(k)}=-\tfrac{1}{2}\mathbf{H}^{-1}\nabla F(\mathbf{u}^{(k)})$
is gradient related, and the standard convergence result for
gradient-related directions with the Armijo
rule~\cite[Sec.~1.2]{bertsekas1999nonlinear} shows that every limit
point is stationary.
\end{proof}

Both assumptions are easy to check: the sensor geometries of
Section~\ref{sec:sim} are not coplanar, and the distances to the
sensors are the denominators already computed in
\eqref{eq:jac_t}--\eqref{eq:jac_r}.
% OPTIONAL (add only after checking your simulation logs):
% ``It was satisfied in all our Monte Carlo runs.''

\subsection{Cramér--Rao Lower Bound}

The Cramér--Rao lower bound (CRLB) bounds the variance of any unbiased estimator of $\mathbf{u}$ from below~\cite{151045}. Under the i.i.d.\ Gaussian noise model of \eqref{eq:meas_model_scalar}, the Fisher information matrix (FIM) for the target position is
\begin{equation}
    \mathbf{J}_F(\mathbf{u}) 
    = \frac{1}{\sigma^2} 
    \sum_{m=1}^{M} \sum_{n=1}^{N}
    (\mathbf{a}_m + \mathbf{b}_n)(\mathbf{a}_m + \mathbf{b}_n)^T,
\end{equation}
where $\mathbf{a}_m = (\mathbf{u} - \mathbf{x}_m^t)/R_m^t$ and $\mathbf{b}_n = (\mathbf{u} - \mathbf{x}_n^r)/R_n^r$ are the unit direction vectors from each transmitter and receiver to the target, respectively. The CRLB on the covariance of any unbiased estimate $\hat{\mathbf{u}}$ satisfies
\begin{equation}
    \mathrm{Cov}(\hat{\mathbf{u}}) \succeq \mathbf{J}_F^{-1}(\mathbf{u}),
\end{equation}
and the scalar root-mean-square error (RMSE) bound is
\begin{equation}
    \mathrm{CRLB} = \sqrt{\mathrm{trace}\!\left(\mathbf{J}_F^{-1}(\mathbf{u})\right)}.
\end{equation}
This bound depends only on sensor geometry and noise variance, and serves as the benchmark in Section~\ref{sec:sim}.

\section{Numerical Results}
\label{sec:sim}

\subsection{Simulation Setup}
We consider a cubic room of side length $100$\,m and place the
$M = N = 4$ transmitting and receiving antennas at its corners. The
transmitter positions (in meters) are
\begin{align*}
    \mathbf{x}_1^t &= [0,\,0,\,0]^\top,     & \mathbf{x}_2^t &= [100,\,0,\,0]^\top,\\
    \mathbf{x}_3^t &= [0,\,100,\,0]^\top,   & \mathbf{x}_4^t &= [100,\,100,\,0]^\top,
\end{align*}
and the receiver positions are
\begin{align*}
    \mathbf{x}_1^r &= [0,\,0,\,100]^\top,   & \mathbf{x}_2^r &= [100,\,0,\,100]^\top,\\
    \mathbf{x}_3^r &= [0,\,100,\,100]^\top, & \mathbf{x}_4^r &= [100,\,100,\,100]^\top.
\end{align*}
Therefore, any target inside the room lies within the convex hull formed by the eight vertices of the cube.
\begin{table}[h]
\centering
\caption{Simulation parameters}
\label{tab:hyperparameters}
\renewcommand{\arraystretch}{1.3}
\begin{tabular}{|l|c|}
\hline
\textbf{Parameter} & \textbf{Value} \\
\hline
Room side length & $100$ m \\
\hline
Noise variance $\sigma^2$ & $5$ dB \\
\hline
Regularization parameter, $\eta$ & $7 \times 10^{-5}$ \\
\hline
Convergence tolerance, $\tau$ & $10^{-8}$ \\
\hline
Backtracking shrink factor, $\beta$ & $0.5$ \\
\hline
Armijo condition constant, $c$ & $10^{-4}$ \\
\hline
$\epsilon_0$ & $10^{-6}$ \\
\hline
\multicolumn{2}{|l|}{\textbf{Baseline geometry (Fig.~\ref{fig:target_location}, Fig.~\ref{fig:rmse_vs_noise})}} \\
\hline
Transmitters, $M$ & $4$ \\
\hline
Receivers, $N$ & $4$ \\
\hline
\multicolumn{2}{|l|}{\textbf{Transmitter sweep (Fig.~\ref{fig:rmse_vs_M})}} \\
\hline
Transmitters, $M$ & $\{4, 8, 12,..., 32\}$ \\
\hline
Receivers, $N$ (fixed) & $4$ \\
\hline
\multicolumn{2}{|l|}{\textbf{Receiver sweep (Fig.~\ref{fig:rmse_vs_N})}} \\
\hline
Transmitters, $M$ (fixed) & $4$ \\
\hline
Receivers, $N$ & $\{4, 8, 12,...,32\}$ \\
\hline
\end{tabular}
\end{table}

The parameters used in the simulation are given in Table~\ref{tab:hyperparameters}.
Performance is compared against the SDP-based method of
\cite{Wang2019} as benchmark. All simulations were run in MATLAB R2023b
on a Windows~11 PC with an Intel Core i7-13700 CPU at 2.10~GHz and
16~GB of RAM.
% TODO: add the solver and timing sentence once confirmed, e.g.
% The SDP benchmark of~\cite{Wang2019} was implemented with CVX~2.2 using
% the SDPT3 solver, and runtimes were measured with \texttt{tic}/\texttt{toc}
% and averaged over 500 Monte Carlo trials.
\begin{figure}[t]
    \centering
    \includegraphics[width=0.8\linewidth]{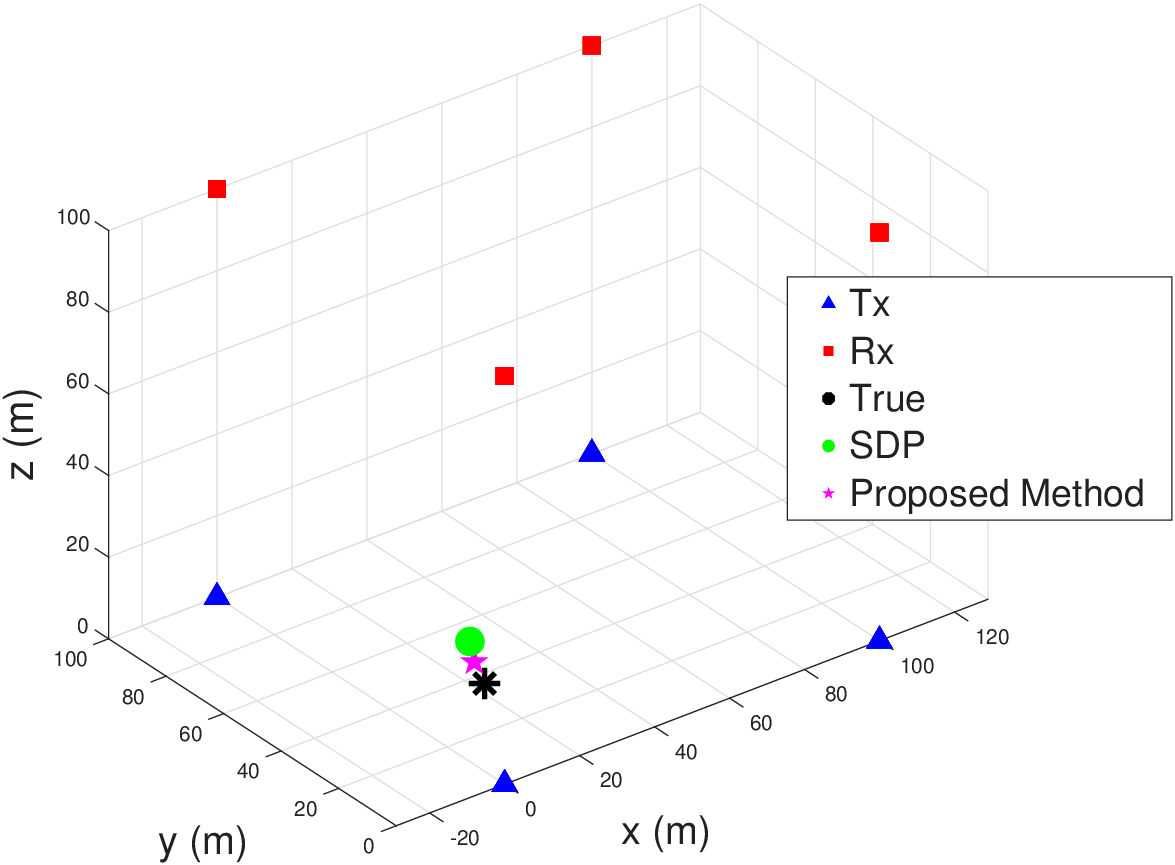}
    \caption{Antenna geometry and target estimates for one noise realization
    with $\sigma^2 = 20$~dB. Transmitters (Tx) are at the bottom corners and
    receivers (Rx) at the top corners of the room, and the true target is at
    $[10,\,20,\,12]^\top$~m. The estimates of SDP~\cite{Wang2019} and the
    proposed DGN method have errors of 16.52~m and 11.49~m, respectively.}
    \label{fig:target_location}
\end{figure}
\begin{figure}[t]
    \centering
    \includegraphics[width=0.8\linewidth]{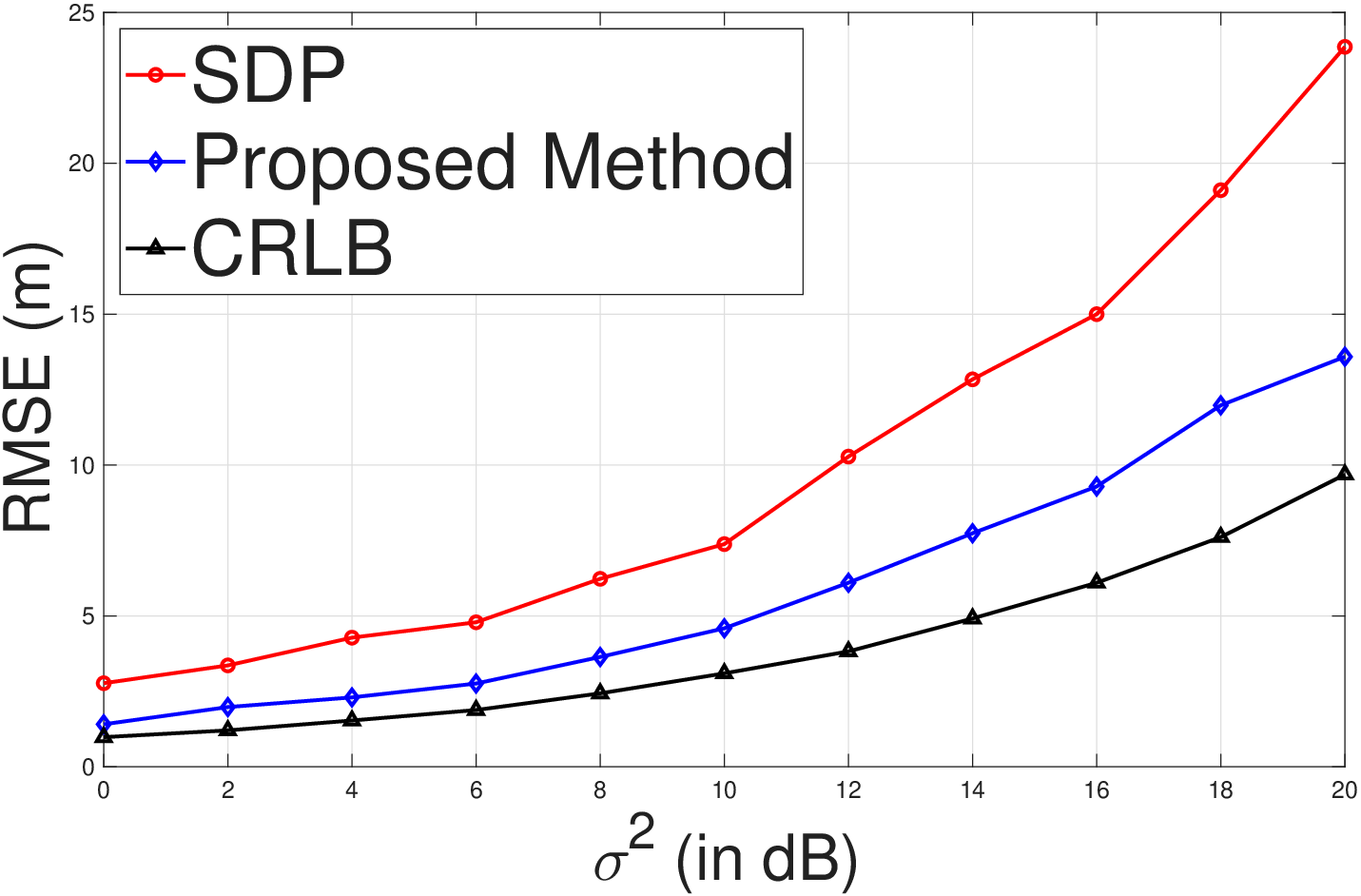}
    \caption{Target localization RMSE vs noise variance ($\sigma^2$)
    for the proposed DGN method, SDP \cite{Wang2019}, and the CRLB.}
    \label{fig:rmse_vs_noise}
\end{figure}

%\subsection{Effect of Noise Power}

In Fig.\ref{fig:target_location}, we are estimating the target whose true location is, $u = [10, 20, 12]^T$ and noise variance is 20 dB, using SDP and our proposed method, where the proposed DGN method estimates the location more accurately (estimation error: 11.4924 m) over SDP method (estimation error: 16.5174 m).

Fig.~\ref{fig:rmse_vs_noise} shows the target localization RMSE as a function of the noise variance $\sigma^2$ for the proposed DGN method, the SDP-based method of \cite{Wang2019}, and the CRLB. The noise variance is swept from 0~dB to 20~dB in steps of 2~dB. As expected, the RMSE of all three curves increases monotonically with $\sigma^2$. The proposed method consistently achieves a lower RMSE than the SDP-based approach across the entire noise range and remains close to the CRLB, demonstrating its superior accuracy and near-optimal performance.

The effect of array size on localization accuracy is examined next by
sweeping $M$ and $N$ independently over a linearly spaced range. Transmitters and receivers are placed on the bottom
($z=0$) and top ($z=100$) planes, respectively, using nested
concentric squares of decreasing half-width, centered at $(50,50,0)$ for transmitter sweep and $(50,50,100)$ for receiver sweep.
To generate an array of size $M$ (respectively $N$), antennas are
placed in groups of four at the corners of concentric squares, so
that $M$ is always a multiple of four. For $K = M/4$ nested squares
indexed by $k = 1,\dots,K$, the half-width of the $k$-th square
decreases linearly from the outermost square (touching the room
boundary) toward the center:
\begin{equation}
    w_k \;=\; 50\left(1 - \frac{k-1}{K}\right), \qquad k = 1,\dots,K,
    \label{eq:half_width}
\end{equation}
and the four transmitter positions contributed by the $k$-th square
are the corners
\begin{equation}
    \mathbf{x}^t \;\in\;
    \left\{(50 \pm w_k,\; 50 \pm w_k,\; 0)\right\},
    \label{eq:square_corners}
\end{equation}
with all four sign combinations included. Receiver positions for the
$N$-sweep are constructed identically on the top plane $z = 100$\,m.

% \begin{table}[h]
% \centering
% \caption{Runtime comparison vs number of transmitter antennas $M$}
% \label{tab:runtime_comparison_M}
% \renewcommand{\arraystretch}{1.3}
% \begin{tabular}{|c|c|c|}
% \hline
% \textbf{M} & \multicolumn{2}{c|}{\textbf{Runtime}} \\
% \cline{2-3}
%  & \textbf{SDP \cite{Wang2019}} & \textbf{Proposed Method} \\
% \hline
% 4  & 0.3244 & 0.0001 \\
% \hline
% 8  & 0.4527 & 0.0002 \\
% \hline
% 12 & 0.5213 & 0.0002 \\
% \hline
% 16 & 0.5790 & 0.0003 \\
% \hline
% 20 & 0.6048 & 0.0003 \\
% \hline
% 24 & 0.6385 & 0.0004 \\
% \hline
% 28 & 0.6721 & 0.0004 \\
% \hline
% 32 & 0.7106 & 0.0005 \\
% \hline
% \end{tabular}
% \end{table}

% \begin{table}[h]
% \centering
% \caption{Runtime comparison vs number of transmitter antennas $M$}
% \label{tab:runtime_comparison_M}
% \renewcommand{\arraystretch}{1.3}
% \begin{tabular}{|c|c|c|}
% \hline
% \textbf{M} & \multicolumn{2}{c|}{\textbf{Runtime}} \\
% \cline{2-3}
%  & \textbf{SDP \cite{Wang2019}} & \textbf{Proposed Method} \\
% \hline
% 4  & 0.2658 & 0.0001 \\
% \hline
% 8  & 0.4954 & 0.0002 \\
% \hline
% 12 & 0.5847 & 0.0002 \\
% \hline
% 16 & 0.3307 & 0.0001 \\
% \hline
% 20 & 0.3061 & 0.0001 \\
% \hline
% 24 & 0.5463 & 0.0003 \\
% \hline
% 28 & 0.5917 & 0.0003 \\
% \hline
% 32 & 0.4450 & 0.0003 \\
% \hline
% \end{tabular}
% \end{table}

\begin{table}[!t]
\centering
\caption{Runtime comparison (in seconds) of SDP~\cite{Wang2019} and the proposed method}
\label{tab:runtime_comparison}
\renewcommand{\arraystretch}{1.3}
\setlength{\tabcolsep}{4pt}
\subfloat[Varying $M$, $N=4$]{%
\begin{tabular}{|c|c|c|}
\hline
$\bm{M}$ & \textbf{SDP} & \textbf{Proposed} \\
\hline
4  & 0.2658 & 0.0001 \\
\hline
8  & 0.3061 & 0.0001 \\
\hline
12 & 0.3307 & 0.0001 \\
\hline
16 & 0.4450 & 0.0002 \\
\hline
20 & 0.4954 & 0.0002 \\
\hline
24 & 0.5463 & 0.0003 \\
\hline
28 & 0.5847 & 0.0003 \\
\hline
32 & 0.5917 & 0.0003 \\
\hline
\end{tabular}}%
\hfil
\subfloat[Varying $N$, $M=4$]{%
\begin{tabular}{|c|c|c|}
\hline
$\bm{N}$ & \textbf{SDP} & \textbf{Proposed} \\
\hline
4  & 0.2897 & 0.0001 \\
\hline
8  & 0.5382 & 0.0002 \\
\hline
12 & 0.5972 & 0.0003 \\
\hline
16 & 0.6017 & 0.0003 \\
\hline
20 & 0.6305 & 0.0004 \\
\hline
24 & 0.6699 & 0.0004 \\
\hline
28 & 0.6936 & 0.0005 \\
\hline
32 & 0.7063 & 0.0005 \\
\hline
\end{tabular}}
\end{table}

\begin{figure}[t]
    \centering
    \includegraphics[width=0.8\linewidth]{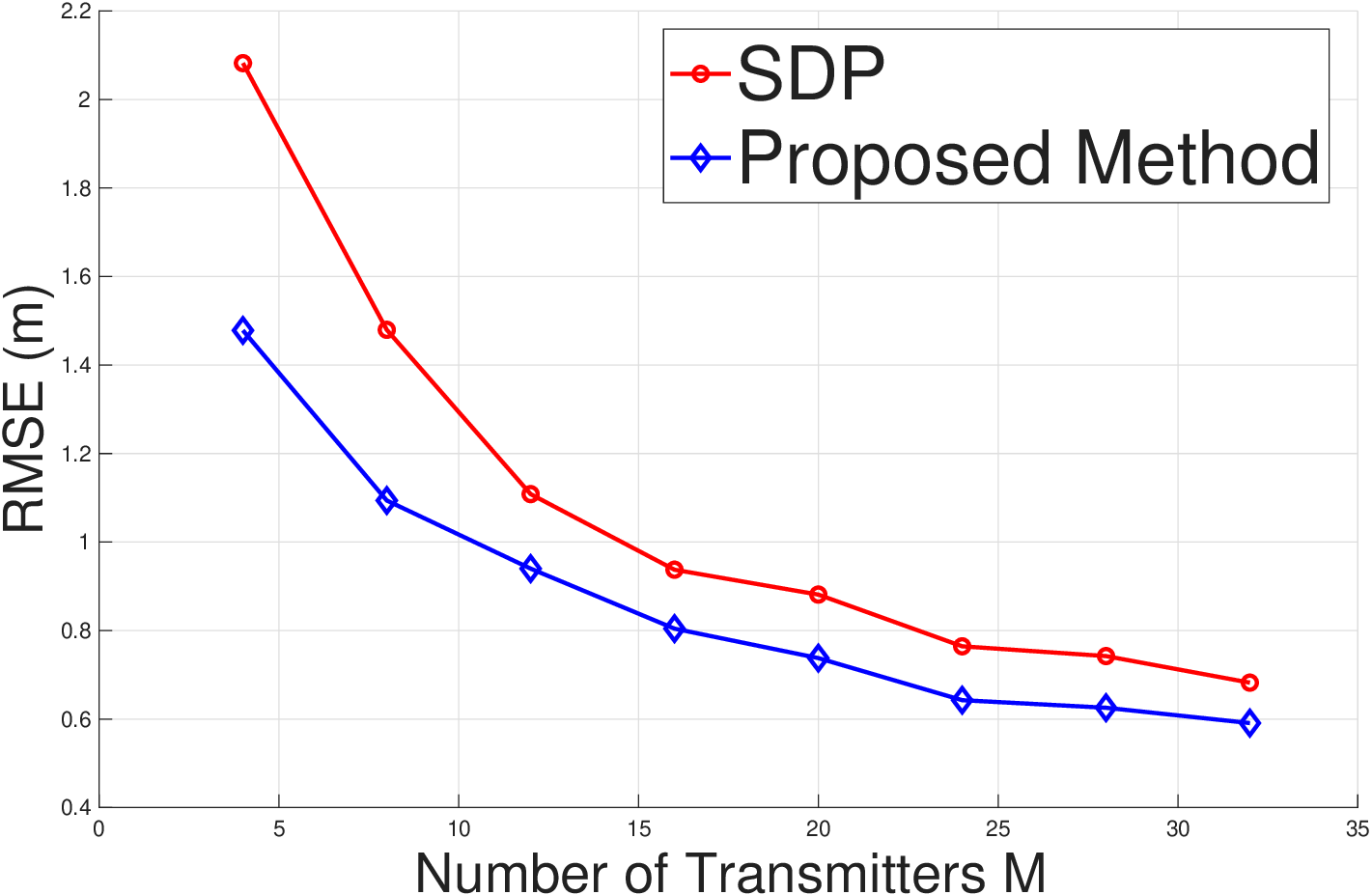}
    \caption{RMSE vs number of transmitting antennas.}
    \label{fig:rmse_vs_M}
\end{figure}
\begin{figure}[t]
    \centering
    \includegraphics[width=0.8\linewidth]{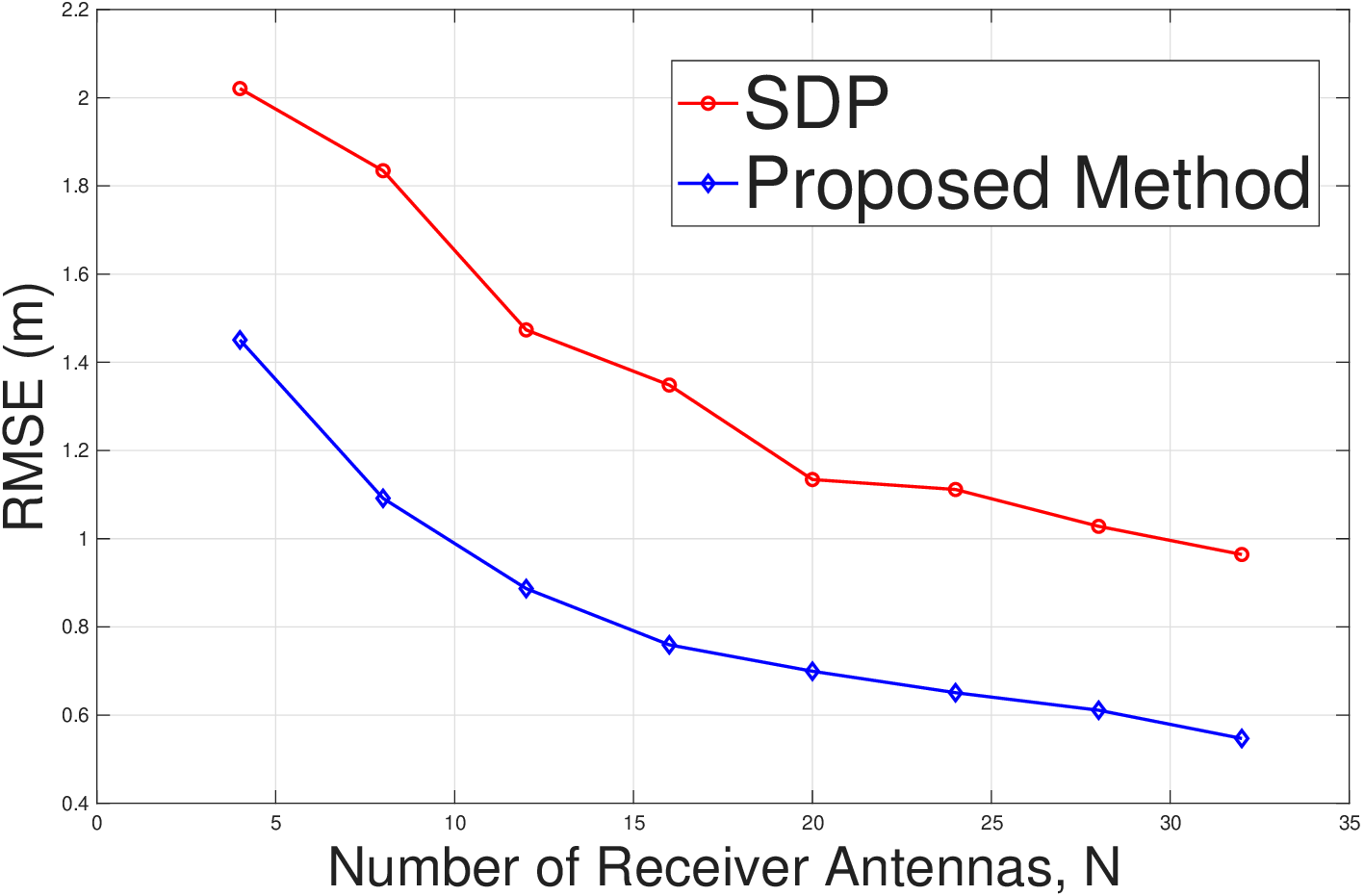}
    \caption{RMSE vs number of receiving antennas.}
    \label{fig:rmse_vs_N}
\end{figure}
%\textcolor{blue}{Figure~\ref{fig:runtime_vs_noise} 
In Fig.~\ref{fig:rmse_vs_M}, the number of receivers is held fixed at
$N = 4$ while the number of transmitters is swept linearly as
$M \in \{4, 8, 12,..., 32\}$, corresponding to $K = 1, 2, 3,..., 8$
nested squares; conversely, in Fig.~\ref{fig:rmse_vs_N}, the number
of transmitters is held fixed at $M = 4$ while the number of
receivers is swept linearly as $N \in \{4, 8, 12,..., 32\}$. Both
figures report the resulting RMSE for the SDP-based method and the
proposed method as a function of $M$ and $N$, respectively. In each
case, the RMSE decreases monotonically with increasing array size,
as the added geometric diversity improves the conditioning of the
localization problem. Across the entire range, the proposed method
consistently achieves lower RMSE than the SDP-based approach, while
also offering substantially lower runtime, as summarized in
Table~\ref{tab:runtime_comparison}(a) for varying $M$ and
Table~\ref{tab:runtime_comparison}(b) for varying $N$.

%% =====================================================================
\section{Conclusion}
\label{sec:conclusion}

In this paper, a damped Gauss--Newton (DGN) algorithm was developed for target localization in MIMO radar systems. The target localization problem was formulated as a regularized nonlinear least-squares optimization problem over the three-dimensional target position. To address the nonconvex nature of the resulting optimization problem, the bistatic range vector was linearized about the current iterate, yielding a Gauss--Newton quadratic model whose minimizer reduces to a closed-form \(3\times3\) linear system, leading to an efficient implementation of the proposed method. An Armijo backtracking line search on the Gauss--Newton direction guarantees monotonic decrease of the objective, and a safeguarded full Gauss--Newton step is additionally taken within each iteration. The proposed DGN method provides an effective and computationally efficient approach to target localization in MIMO radar systems. The results demonstrate its potential for achieving accurate localization while maintaining low computational complexity. Future work will extend the proposed method to the multi-user case, in which several targets must be localized jointly and each bistatic range measurement must first be associated with the correct target. It will also study the performance in larger indoor areas and the minimum number of transmit and receive antennas required to achieve a given localization accuracy.
\bibliographystyle{IEEEtran}
\bibliography{ref}

@INPROCEEDINGS{Wang2019,
  author={Wang, Hongyi and Zhang, Bingbing and Zheng, Liming and Wu, Jianfei},
  booktitle={2019 27th European Signal Processing Conference (EUSIPCO)}, 
  title={{Semidefinite Programming for MIMO Radar Target Localization using Bistatic Range Measurements}}, 
  year={2019},
  volume={},
  number={},
  pages={1-5},
  doi={10.23919/EUSIPCO.2019.8903059}}

@book{richards2005fundamentals,
  title={{Fundamentals of Radar Signal Processing}},
  author={Richards, M.A.},
  isbn={9780071444743},
  lccn={2005047891},
  series={  Professional Engineering},
  url={https://books.google.co.in/books?id=VOvsJ7G1oDEC},
  year={2005},
  publisher={Mcgraw-hill}
}

@INPROCEEDINGS{Guo2019,
  author={{Guo, Yi-duo and Zhang, Yong-shun and Ning-ning Tong and Di Shen}},
  booktitle={Proceedings of 2011 IEEE CIE International Conference on Radar}, 
  title={{Spatial localization of quasi-stationary targets using bistatic MIMO radar}}, 
  year={2011},
  volume={2},
  number={},
  pages={1788-1791},
  doi={10.1109/CIE-Radar.2011.6159918}}

@ARTICLE{Amiri2017,
  author={Amiri, Rouhollah and Behnia, Fereidoon},
  journal={IEEE Signal Processing Letters}, 
  title={{An Efficient Weighted Least Squares Estimator for Elliptic Localization in Distributed MIMO Radars}}, 
  year={2017},
  volume={24},
  number={6},
  pages={902-906},
  doi={10.1109/LSP.2017.2697500}}

@article{Bekkerman2006TargetDA,
  title={{Target Detection and Localization Using MIMO Radars and Sonars}},
  author={Ilya Bekkerman and Joseph Tabrikian},
  journal={IEEE Transactions on Signal Processing},
  year={2006},
  volume={54},
  pages={3873-3883},
 
}

@ARTICLE{5714759,
  author={Xu, Enyang and Ding, Zhi and Dasgupta, Soura},
  journal={IEEE Transactions on Signal Processing}, 
  title={{Source Localization in Wireless Sensor Networks From Signal Time-of-Arrival Measurements}}, 
  year={2011},
  volume={59},
  number={6},
  pages={2887-2897},
  doi={10.1109/TSP.2011.2116012}}

@article{armijo1966minimization,
  title={{Minimization of functions having Lipschitz continuous first partial derivatives}},
  author={Armijo, Larry},
  journal={Pacific Journal of mathematics},
  volume={16},
  number={1},
  pages={1--3},
  year={1966},
  publisher={Mathematical Sciences Publishers}
}

@ARTICLE{Haimovich_2008,
  author={Haimovich, Alexander M. and Blum, Rick S. and Cimini, Leonard J.},
  journal={IEEE Signal Processing Magazine}, 
  title={{MIMO Radar with Widely Separated Antennas}}, 
  year={2008},
  volume={25},
  number={1},
  pages={116-129},
  doi={10.1109/MSP.2008.4408448}}

@ARTICLE{beck_2008,
  author={Beck, Amir and Stoica, Petre and Li, Jian},
  journal={IEEE Transactions on Signal Processing}, 
  title={{Exact and Approximate Solutions of Source Localization Problems}}, 
  year={2008},
  volume={56},
  number={5},
  pages={1770-1778},
  doi={10.1109/TSP.2007.909342}}

@article{Golub1999TikhonovRA,
  title={{Tikhonov Regularization and Total Least Squares}},
  author={Gene H. Golub and Per Christian Hansen and Dianne P. O’Leary},
  journal={SIAM J. Matrix Anal. Appl.},
  year={1999},
  volume={21},
  pages={185-194},
}

@book{151045,
author = {Kay, Steven M.},
title = {Fundamentals of statistical signal processing: estimation theory},
year = {1993},
isbn = {0133457117},
publisher = {Prentice-Hall, Inc.},
address = {USA}
}

@book{nocedal2006numerical,
  author    = {Nocedal, Jorge and Wright, Stephen J.},
  title     = {Numerical Optimization},
  edition   = {2nd},
  publisher = {Springer},
  address   = {New York, NY, USA},
  year      = {2006}
}

@book{bertsekas1999nonlinear,
  title={Nonlinear Programming},
  author={Bertsekas, D.P.},
  isbn={9781886529007},
  lccn={99732088},
  series={Athena scientific optimization and computation series},
  year={1999},
  publisher={Athena Scientific}
}

\end{document}